# Finding the Inner Clock:
# A Chronobiology-based Calendar

**Sarah Janböcke**
University of Siegen
Siegen, 57072, Germany
sarah.janboecke@uni-siegen.de

**Alina Gawlitta**
University of Siegen
Siegen, 57072, Germany
alina.gawlitta@gmail.com

**Judith Dörrenbächer**
University of Siegen
Siegen, 57072, Germany
judith.doerrenbaecher@uni-siegen.de

**Marc Hassenzahl**
University of Siegen
Siegen, 57072, Germany
marc.hassenzahl@uni-siegen.de

## Abstract
Time and its lack of play a central role in our everyday lives. Despite increasing productivity, many people experience "time stress" – exhaustion and a longing for time affluence, and at the same time, a fear of not being busy enough. All this leads to a neglect of natural time, especially the patterns and rhythms created by physiological processes, subsumed under the heading of chronobiology. The present paper presents and evaluates a calendar application, which uses chronobiological knowledge to support people's planning activities. Participants found our calendar to be interesting and engaging. It especially made them think more about their bodies and appropriate times for particular activities. All in all, it supported participants in negotiating external demands and personal health and wellbeing. This shows that technology does not necessarily has to be neutral or even further current (mal-)practices. Our calendar cares about changing perspectives and thus about enhancing users' wellbeing.

## Author Keywords
Wellbeing; health; calendar; chronobiology; time management; work-life-balance; digital calendar

## CSS Concepts
● Human-centered computing~Human computer interaction (HCI)  • Human-centered computing~User studies  • Human-centered computing~Field studies
https://dl.acm.org/ccs/ccs_flat.cfm

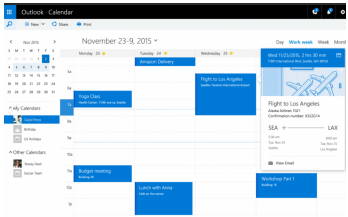

**Figure 1:** Outlook Calendar

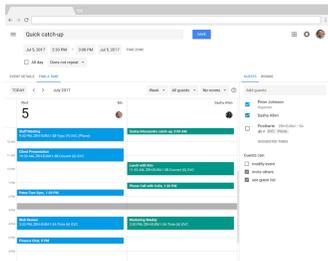

**Figure 2:** Google Calendar

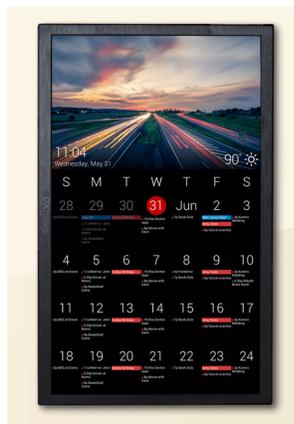

**Figure 3:** DAKBoard

### Time, calendars and health

When we consider time as a phenomenon, we often refer to two quite different views: natural time and mechanical time [1]. Natural time is circular, a matter of natural rhythms, such as ebb and flow, day and night, or the seasons. Even our bodies, down to basal cell mechanisms, follow a number of circular patterns and rhythms [7]. In contrast, mechanical time is linear and standardized. Represented by clocks and calendars, it is crucial to an industrialized society. It organises schedules and allows for planning.

Given its close relationship to economic processes and its focus on efficiency, society and especially work stresses mechanical time [4,7,14–16]. People respond with "time stress" – on the one hand, they long for time affluence [12]. The need to be always-on and always-connected creates a feeling of exhaustion [15]. On the other hand, "busyness" becomes a way to create a positive identity as being valuable and needed [14].

When the feeling of lacking time mounts, people tend to optimise themselves [14]. Calendars are crucial to this. They help to organize the own schedule and activities, but also contain important social and emotional aspects. Calendar use can be related to the experience of pride in one's own achievements, as well as the feeling of having control over one's own life. To do so, calendars have to be able to record different scenarios and to reproduce them in a visually appealing way. For this reason, there are systems for a wide variety of needs such as the Asana Calendar, the MS Office products or HCL Notes for work contexts. Within the family context joint planners form an opportunity to share related activities, e.g. Cozi Family Organizer, DAKBoard or MyCalou (Figure 1-3). Some calendar systems cover special target groups, e.g. TheMoon [1,10,18,21].

While calendars provide structure to organize everyday tasks, they do so without any concerns about their users' wellbeing or healthy time use (see [8,17] for exceptions). They rather appear as presumably neutral technologies to support whatever schedules their users prefer.

### Wellbeing and chronobiology

As already pointed out, natural time, its rhythms and patterns, is deeply ingrained into our bodies and the environment. Mechanical time does rarely acknowledge this. Health and wellbeing, though, may be dependent on the appropriate phasing of inner clocks with "external" events.

Chronobiology *"is the science of investigating and objectively quantifying phenomena and mechanisms of the biologic time structure, including the rhythmic manifestations of life."* [18, Online Glossary]. Chronobiology points out that bodily processes in humans are repeated in intervals within 24 hours. Thus, the releases of melatonin, serotonin, glutamates or endorphins in different rhythms are naturally controlled. Due to the body's hormone concentrations certain times of the day fit better to particular activities than others [2,5–7,22]. The internal intervals vary from person to person. However, a standardization can be derived from sleep behaviour in accordance to the Munich Chronotype Questionnaire (MCQ) [11].

In general, two different, so-called chronotypes can be defined. Larks are characterised by early awakening. They reach their mental and physical peaks within the first half of the day. Owls are the opposite. They raise late and reach their mental and physical peaks later within the day. In accordance to the MCQ it is possible to create a rough process plan of peoples' internal intervals to support the adaption of activities to the indi-vidual chronotype and natural rhythm. In other words, we can define suitable times for relaxation due to natural hormonal balances, as well as optimal phases for problem-solving events, inspirational tasks and regeneration phases.

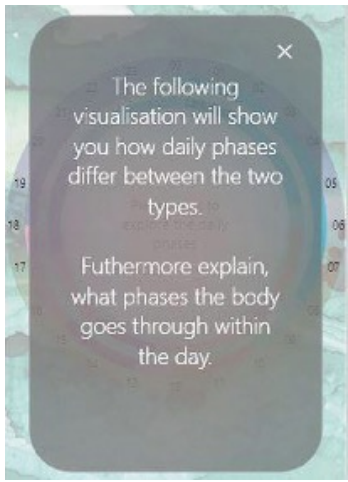

**Figure 4**: Section of click dummy- Knowledge transfer I

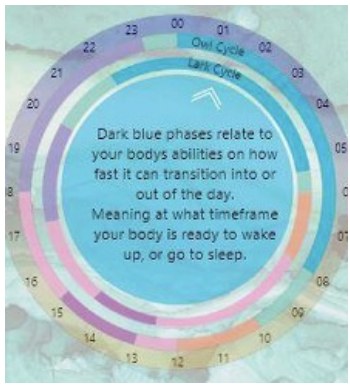

**Figure 5**: Section of click dummy- Knowledge transfer II

Disregarding one's internal clock results in sleep disturbances, a weakened immune system and thus a higher susceptibility to disease and according negative effects on personal wellbeing. What is needed is a way to structure the day not only in a productive but also natural way [11]. By merging chronobiological insights with a calendar, we become able to communicate suitable times for particular activities and thereby improve wellbeing.

In the remainder of the paper, we present a calendar application, which supports its users with reconciling external demands with their inner clock.

## A calendar based on chronobiology

Our research focused on the question of whether a digital calendar can influence peoples' wellbeing by nudging them into "healthy" time-use based on chronobiological insights [7]. We began with an online survey (n=33), followed by a concept and prototype creation phase and finished with two user studies; a laboratory study (n=10) and a field study (n=5).

*Survey*
The survey (n=33) was conducted anonymously-online and was disseminated via social media channels. Four participants of the study classified themselves as pupils, 22 as students in university, six as employees and one as freelancer Twelve percent were younger than 18 years, 58% = 19-25 years, 24% = 26-35 years, 3% = 36-49 years and 3% over 50 years. The questionnaire contained six questions that focused on the active use of calendars (e.g. daily use, one or more calendars, typical categories).

The survey showed that the vast majority of participants use digital calendars to help them structure their private and professional daily lives jointly (25 of 33; 76%). Category-wise birthdays (30 of 33; 91%), private appointments, such as medical ones (28 of 33; 85%), work-related appointments (23 of 33; 70%) and family-related meetings (24 of 33; 73%) carry the volume of planning. Further typical events are appointments with friends (22 of 33; 67%), sporting activities (15 of 33; 46%) and relaxation activities, including activities related to wellbeing, such as meditation, time for oneself or yoga (10 of 33; 30%). Five participants (5 of 33; 15%) stated that they enter routine activities, such as getting up, sleeping or meal times to their calendars. The survey's results state a clear preference among the participants for just using one digital calendar. Six categories of events can be distinguished: work, friends, family, sports, mindfulness and other activities, such as general private appointments. The survey's findings were incorporated in our following concept.

*Concept and prototype*
The concept aims at transferring knowledge about chronobiologic processes to daily planning and visually supporting the planning from a wellbeing perspective (Fig. 4). First, the user is introduced to knowledge about chronobiological phases and the calendar's "intention", to plan activities in line with chronobiology, if possible (Fig.5). We then administered a questionnaire (based on the MCQ [11]) to classify the user either into the "lark" or "owl" types. To enter an event, users choose a date and then select an event type from the six categories derived from the survey (i.e., work, friends, family, sports, mindfulness, and other activities) (Fig. 6). Subsequently, the user is asked to fill in detail information as in any other calendar application. Entries are visually supported by showing immediately, whether the activity planned fits into a harmonious day cycle. If the planned appointments do not match chronobiological needs, they are visualised diagonally and colourlessly. In addition, these appointments "wobble" within the calendar to create friction and to seek user awareness (Fig. 7). Thus, by intuitive nudges users are alerted about chronobiologically "mis-planned" events. This visual feedback seeks constant attention and is intended to trigger active reflection (Fig. 8) [9]. However, it should be possible to plan the appointments despite inappropriate time, without hindering the user's planning process [13,20]. There are many external reasons, why an appointment needs to be accepted, which may not be in line with one's chronobiological demands. While the user can freely choose, the arrangement of the activities and the colour scheme provide impulses for more "appropriate decisions". This heightened presence of chronobiological knowledge is meant to support the development of "time-competence".

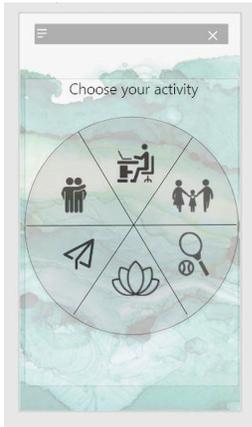

**Figure 6**: Click dummy- Activity selection

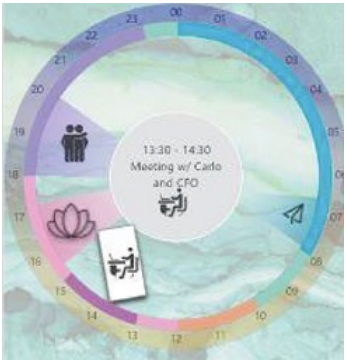

**Figure 7**: Section of click dummy-Visualisation of mis-planned events

*Study 1- laboratory study*
We carried out a lab study (n=10, age range = 23 to 42, female = 6, male = 4, M = 28.4) with a click-dummy based on our concept and the base functions mentioned above. Both user studies were accompanied by semi-structured interviews (before and after use) that should provide insight into user experience and app influence on daily habits.

In our lab study participants were approached to briefly test the click-dummy app and asked to plan a few appointments. They were then asked for their opinion about the process and whether they could imagine our calendar for daily use.

Eight out of ten participants had particularly positive emotions, when using the application. They found the topic of the calendar to be interesting. Two participants expressed scepticism. They felt that the time span of testing was too short to state a real opinion. The participants expressed that the application was intuitive, informing, enjoyable and that they could imagine benefits in everyday life.

Reasons why the application would be beneficial for users included the establishment of a deeper understanding of the body and its hormones. It was mentioned that the focus on health and the body could be brought back into daily life. Five participants stated that they could imagine using the calendar on a daily basis, but a prerequisite would be a mobile app. One participant also mentioned that the calendar is well usable for private use, but not in professional life and must be adapted to be so.

*Study 2 - field study*
Based on the click dummy, a simplified dynamic application was built to actually use and test the calendar in the field (Fig.9). The basic structure of the calendar is built in HTML and PHP with the use of MySQL. For the field study, the application was installed together with the participants onto their personal laptops. The participants were asked to use the chronobiologic calendar as their main calendar for seven days. Five people participated (n=5, age range = 22 to 55, female = 3, male = 2, M = 31.6).

Before the study, participants were asked four questions, aimed at finding how satisfied participants feel with their general daily planning, which emotions participants have in relation to their planner, whether the participants felt sufficiently supported by their planner and what wishes they had for further support. The participants were very satisfied with their original calendar usage and did not express any need for change. After the 7-day usage participants were interviewed again.

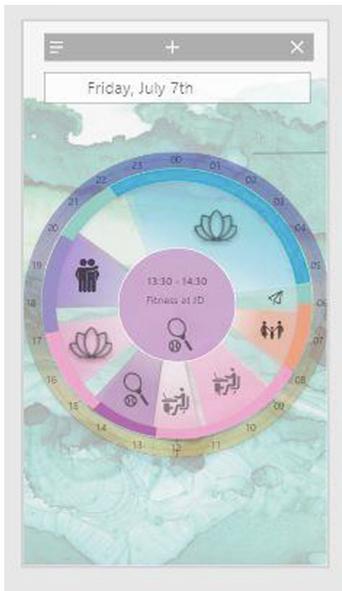

**Figure 8**: Click dummy - Visualisation of a perfect day

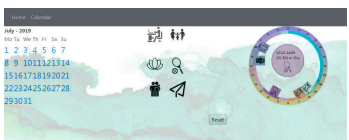

**Figure 9**: Calendar – dynamic laptop application

In general, the use of our calendar was found to be pleasant and even fun, but the new routine irritated at first: "I was confused by the calendar telling me my appointments are at wrong places. It took me one or two days to really get into the idea of placing appointments into the right phases and how they fit the best."

By using the laptop, planning took more time than usual. After the initial familiarisation, the planning technique shifted from constantly opening the calendar on the laptop, to a calendar's one-time use at the beginning of the day. Nevertheless, the use was described as pleasant. P4 described the bridging via laptop as "little bit of an adventure". Especially, the visual confirmation of how "well" the day was planned was perceived positively: "I never ever thought about the fact that every activity has a certain time where it is optimal to perform it. That thought really got me interested into the topic."

None of the participants had any problems understanding the application. Yet, all participants used their regular calendar in parallel: "Like I said I could not take care of the bad timings all the time, which was ok to me at first. But after a little it started to bug me, especially when all the other activities had a good fit. It was opening up a challenge for me to re-check." Some participants expressed difficulties with the category system of the calendar. The desire for individual freedom was mentioned, such as having more options for colour design, lists or overviews: "I need more ways to make the calendar mine."

Our main question aimed at the participants' wellbeing and whether they felt more positive or relaxed. P1 said, for example, that he felt positively confirmed by the use of the calendar in his daily planning. For P2, an overall raised mood was not clearly identifiable, as the time span was too short for such a statement, but the feeling during calendar use was clearly positive. Also, P3, P4 and P5 had positive feelings during and by using the calendar. P4 even said that she actually felt more positive and relaxed and also emphasised that the understanding of her own body had improved. P4 would even like to further integrate this new knowledge into her live. For P5 the test phase gave positive impulses, even if the stress level had not decreased: "I plan my time more carefully or with more perspective on how much time I put into my work-life-balance. I actually stopped working after my highly-effective phase because I felt my brain shutting down."

However, it should be stressed that also negative experiences emerged. P2 expressed that the activity evaluations triggered negative feelings: "(...) I felt upset scheduling "bad" times. So, when I knew I had to plan "poorly", I really didn't want to schedule (...). I got used to it after a day or two but at first it was leading me to thoughts like: This appointment will be bad, because I am not in the right mood."

## Discussion

The two studies showed that the calendar was positively accepted by the majority of the participants. The calendar provided reason, backed up with knowledge, to get time for oneself and not to feel guilty about it. This is especially interesting since the participants apparently felt pressured to fill their calendar with work activities, which supports the notion of the stigma of "too much" leisure time. By using the chronobiologic calendar, the participants felt less pressured to insert further working hours, but instead, thought more about time for relaxation, friends or family, which presumably will lead to more wellbeing. The calendar's design triggered reflection regarding "unbalanced" planning and even provoked a challenge to create a perfect day, productive but in line with natural rhythms. We argue, that through technologies, such as this calendar, the social stigma of always being busy and hardworking, could be attenuated. In this context, technology suffers from a bad reputation, emphasizing speed, efficiency, and self-exploitation [3]. However, whether technology has these effects, depends on its design. To the contrary, the present study shows, that technology can be used to slow down time, i.e., to make a more reflective use of it.

Future studies will consider demands for further individualisation with regard to colours, design/GUI and interaction. Since the participants were not willing to exchange their main calendars with our chronobiologic calendar, we will look into strategies how to integrate the tested app into existing calendar ecologies and, thus, add value to existing apps with a wellbeing focus. This would give opportunity to explore how conflicts between healthy time usage and external, work-related demands could be negotiated with the support of technology.

The response to the general idea of our concept was particularly positive. This shows, that technology can not only be used to further prolong existing (mal-)practices, but also to create reflection and more healthy practices. While the market for mindfulness applications in general is booming, the present strategy is different. We rather focus on presumably "neutral" everyday technologies, such as calendars, and their potential to reframe existing practices in line with better and healthier living.